\documentclass[aps,prd,showpacs,superscriptaddress,nofootinbib,preprintnumbers,notitlepage]{revtex4-1} 
\usepackage{graphicx}
\usepackage{epsf}
\usepackage{bm}
\usepackage{amsmath}
\usepackage{amsfonts}
\usepackage{amssymb}
\usepackage{epstopdf}
\usepackage{natbib}
\usepackage{hyperref}
\usepackage{color}
\usepackage{verbatim}
\usepackage{multirow}
\usepackage{bm}
\usepackage{hyperref}
\usepackage[normalem]{ulem}

\definecolor{darkblue}{rgb}{0.0, 0.0, 0.55}
\definecolor{darkred}{rgb}{0.55, 0.0, 0.0}
\usepackage{hyperref}
\hypersetup{
    colorlinks=true,
    linkcolor=darkblue,
    citecolor=darkblue,
    urlcolor=darkblue}
    
\makeatletter\let\expandableinput\@@input\makeatother

\begin{document}

\title{Linear Growth of Matter Perturbations Probed by Redshift-Space Distortions in Interacting $\Lambda(t)$CDM Cosmologies}
%\title{Observational Constraints on $S_8$ from Newtonian Perturbations of $\Lambda(t)$CDM Models}

\author{A. A. Escobal}
\email{anderson.aescobal@outlook.com}
\affiliation{Department of Astronomy, University of Science and Technology of China, Hefei, Anhui 230026, China}
\affiliation{School of Astronomy and Space Science, University of Science and Technology of China, Hefei, Anhui 230026, China.}
\affiliation{Universidade Estadual Paulista (UNESP), Faculdade de Engenharia e Ci\^encias, 
Departamento de F\'isica, Av. Dr. Ariberto Pereira da Cunha 333, 
12516-410, Guaratinguet\'a, SP, Brazil}

\author{H. A. P. Macedo}
\email{ha.macedo@unesp.br}
\affiliation{Universidade Estadual Paulista (UNESP), Faculdade de Engenharia e Ci\^encias, 
Departamento de F\'isica, Av. Dr. Ariberto Pereira da Cunha 333, 
12516-410, Guaratinguet\'a, SP, Brazil}

\author{J. F. Jesus}
\email{jf.jesus@unesp.br}
\affiliation{Universidade Estadual Paulista (UNESP), Faculdade de Engenharia e Ci\^encias, 
Departamento de F\'isica, Av. Dr. Ariberto Pereira da Cunha 333, 
12516-410, Guaratinguet\'a, SP, Brazil}
\affiliation{Universidade Estadual Paulista (UNESP), Instituto de Ciências e Engenharia, 
Departamento de Ci\^encias e Tecnologia, R. Geraldo Alckmin 519, 
18409-010, Itapeva, SP, Brazil}

\author{R. C. Nunes}
\email{rafadcnunes@gmail.com}
\affiliation{Instituto de F\'isica, Universidade Federal do Rio Grande do Sul, 
91501-970 Porto Alegre RS, Brazil}
\affiliation{Divis\~ao de Astrof\'isica, Instituto Nacional de Pesquisas Espaciais, 
Avenida dos Astronautas 1758, 
12227-010, S\~ao Jos\'e dos Campos, SP, Brazil}

\author{J. A. S. Lima}
\email{jas.lima@iag.usp.br}
\affiliation{IAG, Universidade de S\~ao Paulo, 05508-900 S\~ao Paulo, SP, Brazil}

\begin{abstract}
In the context of a spatially flat $\Lambda(t)$CDM cosmology, we investigate interacting dark energy (IDE) scenarios characterized by phenomenological interaction terms proportional to the Hubble expansion rate and the dark energy density. Our analysis is performed at both the background and linear perturbation levels, with particular emphasis on the evolution of dark matter density fluctuations. Cosmological constraints are derived from a joint analysis of CMB distance priors, Baryon Acoustic Oscillations (BAO), Type Ia supernovae (SNe Ia) from Pantheon+, Redshift-Space Distortions (RSD), and $H(z)$ data from Cosmic Chronometers (CC). Using the linear growth of matter perturbations, we estimate the clustering parameter $S_8$ within IDE extensions of the flat $\Lambda(t)$CDM framework. At the perturbative level, we consider interaction terms of the form $Q_{\text{I}}=\varepsilon a H\bar{\rho}_{\Lambda(t)}$ (Model I) and $Q_{\text{II}}=\varepsilon H\bar{\rho}_{\Lambda(t)}$ (Model II). From the combined dataset, we obtain the constraints $S_8 = 0.870 \pm 0.026$ for Model I and $S_8 = 0.872 \pm 0.026$ for Model II. Finally, we discuss the implications for the coupling parameter $\varepsilon$, taking into account the semi-analytical approximations and observational data employed in this study.
\end{abstract}
%Despite its remarkable empirical success, the ΛCDM framework faces several long-standing and newly emerging challenges. On the theoretical side, the model is affected by the cosmological constant and coincidence problems. On the observational side, the past decade has revealed statistically significant tensions and anomalies, most prominently the Hubble constant (H₀) tension and the S₈ tension.

\maketitle

\section{Introduction}

The $\Lambda$ Cold Dark Matter ($\Lambda$CDM) model constitutes the standard paradigm of modern cosmology. It successfully accounts for a wide range of astronomical observations, spanning from the early Universe—such as primordial nucleosynthesis \cite{ParticleDataGroup:2022pth} and the Cosmic Microwave Background (CMB) anisotropies \cite{Planck2018}—to late-time probes, including Type Ia supernovae (SNe Ia) \cite{Brout:2022vxf}, cosmic chronometer measurements of the expansion rate $H(z)$ \cite{Moresco:2022phi}, and the formation of large-scale structure. Despite this remarkable success, the $\Lambda$CDM framework faces {long-standing and newly emerging} challenges. On the theoretical front, for instance, the model is plagued by the cosmological constant and Coincidence problems \cite{Weinberg:1988cp, Carvalho:1991ut, Lima:1994gi, Zlatev:1998tr,Perico:2013mna,Lima:2013dmf}. More recently,  a number of statistically significant observational tensions and anomalies have emerged over the past decade, most notably the Hubble constant ($H_0$) tension and the $S_8$ tension. These discrepancies, observed between early - and late-Universe measurements, may point to unresolved systematic effects or, more intriguingly, to physics beyond the standard cosmological model. For a comprehensive and up-to-date review of these issues, see Ref.~\cite{CosmoVerseNetwork:2025alb}.

A particularly noteworthy issue concerns the emerging tension associated with the parameter
\begin{equation}
S_8 \equiv \sigma_8 \sqrt{\Omega_{\rm m}/0.3}, 
\end{equation}
which characterizes the normalization of matter density fluctuations on
$8\,h^{-1}\,\mathrm{Mpc}$ scales. Recent weak-lensing analyses, most notably those from the KiDS-1000 collaboration~\cite{Wright:2025xka}, exhibit a level of consistency with the Planck CMB constraints. In contrast, full-shape galaxy clustering studies continue to indicate a pronounced mismatch with early-Universe predictions, with reported discrepancies reaching up to $4.5\sigma$~\cite{Ivanov:2024xgb,Chen:2024vuf}. 

Independent indications of deviations in $S_8$ are also found in redshift-space distortion measurements~\cite{Nunes:2021ipq, Kazantzidis:2018rnb, Benisty:2020kdt} and in several other late-time observational probes~\cite{Karim:2024luk,Dalal:2023olq}. Collectively, these results—derived from distinct and complementary cosmological data sets, have fueled sustained interest in extensions of the standard cosmological framework, potentially signaling the need for new physics beyond the $\Lambda$CDM model (see Ref.~\cite{CosmoVerseNetwork:2025alb,Perivolaropoulos:2021jda} for an overview).

Interacting dark energy (IDE) frameworks have recently gained renewed attention in the literature as natural candidates for alleviating the growing set of cosmological tensions. Although such models are rooted in a long-standing theoretical tradition and have been motivated in cosmology for several decades (see~\cite{vanderWesthuizen:2025rip,Bolotin:2013jpa,Wang:2024vmw} for a comprehensive review), they have experienced a resurgence of interest in light of recent observational discrepancies. In particular, IDE scenarios have been extensively explored as possible resolutions to the Hubble constant ($H_0$) tension~\cite{DiValentino:2017iww, Kumar:2017dnp, Vagnozzi:2019ezj, Zhang:2025dwu, Patil:2023rqy, Yang:2022csz,Yao:2022kub,Amiri:2021kpp,Montani:2024pou,Zhang:2025dwu,Pan:2020bur,Akarsu:2025gwi,Yang:2020uga,Kumar:2019wfs, DiValentino:2019ffd, Brito:2024bhh,Escobal:2026zxb}, as well as in attempts to address the $S_8$ tension~\cite{Sabogal:2024yha,Lucca:2021dxo,Yashiki:2025loj,Ghedini:2024mdu,Shah:2024rme}. While these models provide a flexible and well-motivated framework capable of improving the fit to individual data sets—and in some cases outperforming the $\Lambda$CDM model~\cite{Silva:2025hxw,Giare:2024smz,Li:2026xaz,Figueruelo:2026eis,Paliathanasis:2026ymi,Pan:2025qwy,Zhai:2025hfi,Chakraborty:2025syu,Li:2025owk,Li:2024qso,Li:2025muv,You:2025uon,Chakraborty:2025syu,Yang:2021oxc}—achieving a simultaneous and consistent resolution of both tensions remains a significant theoretical and phenomenological challenge.

In this work, we investigate the possibility of interactions within the dark sector by means of an IDE model, with the specific aim of alleviating the $S_8$ tension in the context of redshift-space distortion (RSD) data. The general formulation governing the evolution of scalar perturbations in IDE scenarios is well established in the literature (see, e.g.,~\cite{Gavela:2010tm,Valiviita:2008iv,Yang:2018euj,Majerotto:2009np,Johnson:2020gzn}). In addition, simplified treatments of linear perturbations in the quasi-static and sub-horizon regimes, applicable to the IDE subclasses considered in this work, have been extensively investigated in Refs.~\cite{LimaEtAl97,Jesus:2011ek,DelPopolo:2012dq}. Recent progress toward modeling IDE scenarios at the non-linear level has been presented in Refs.~\cite{Silva:2025bnn,Tudes:2024jpg}.

We aim to place observational constraints on the model parameters, most notably $S_8$, by employing a comprehensive and diverse data set. Our analysis includes RSD measurements compiled in Ref.~\cite{Avila:2022xad}, cosmic chronometers (CCs) \cite{Moresco:2022phi}, and the latest baryon acoustic oscillation (BAO) measurements from DESI DR2 \cite{DESI:2025zgx}. In addition, we incorporate three independent Type Ia supernova (SNe Ia) compilations—Pantheon+ \cite{Brout:2022vxf}, DESY5 \cite{DES:2024jxu}, and Union3 \cite{Rubin:2023jdq}—together with CMB distance priors derived from Planck 2018 \cite{Chen:2018dbv}.

This article is organized as follows. In Section~\ref{FIDE}, we present the dynamical formulation of interacting dark energy models and devote that section to the study of linear density perturbations in the sub-horizon regime, which is then applied to two specific IDE scenarios. In Section~\ref{data}, we describe the observational data sets and the statistical methodology adopted in our analysis. Our results and their implications are discussed in Section~\ref{Res}. Finally, we summarize our main conclusions and outline future perspectives in Section~\ref{Con}.

\section{Semi-analytical treatment in the sub-horizon regime}
\label{FIDE}

The assumption of a non-gravitational interaction between dark energy and dark matter leads to a coupled evolution of their energy densities. This interaction can be phenomenologically described by introducing a source term $Q$ in the conservation equations, such that the energy--momentum tensors of the dark sector components are no longer conserved individually but satisfy
\begin{align}
\nabla_\mu T^{\mu\nu}_{\rm DM} &= Q^\nu \,, \\
\nabla_\mu T^{\mu\nu}_{\rm DE} &= -Q^\nu \,,
\end{align}
ensuring the total conservation of energy--momentum,
$\nabla_\mu\left(T^{\mu\nu}_{\rm DM}+T^{\mu\nu}_{\rm DE}\right)=0$.
Here, $Q^\nu \equiv (Q,0,0,0)$ denotes the energy--momentum transfer four-vector in the comoving frame, implying energy exchange without momentum transfer at the background level.

Within a spatially flat Friedmann--Lemaître--Robertson--Walker (FLRW) spacetime, the background expansion dynamics are governed by the Friedmann equation
\begin{equation}
H^2 = \frac{\kappa^2}{3}
\left(
\bar\rho_b + \bar\rho_r + \bar\rho_{\rm DM} + \bar\rho_{\rm DE}
\right),
\label{H2}
\end{equation}
where $\kappa^2 \equiv 8\pi G$, and an overbar denotes background quantities.

The evolution of the individual energy densities is determined by the continuity equations
\begin{align}
\dot{\bar\rho}_b + 3H\bar\rho_b &= 0 \,, \label{b} \\
\dot{\bar\rho}_r + 4H\bar\rho_r &= 0 \,, \label{rad} \\
\dot{\bar\rho}_{\rm DM} + 3H\bar\rho_{\rm DM} &= Q \,, \label{DMQ} \\
\dot{\bar\rho}_{\rm DE} &= -Q \,, \label{DEQ}
\end{align}
where an overdot denotes differentiation with respect to cosmic time.

In this framework, baryons and dark matter are assumed to be pressureless fluids, $p_b=p_{\rm DM}=0$, while radiation satisfies the relativistic equation of state $p_r = \bar\rho_r/3$. The dark energy component is modeled as a vacuum fluid with equation-of-state parameter $w_{\rm DE}=-1$, \emph{characterizing the class of $\Lambda(t)$CDM models}. As a result, the interaction term $Q$ controls the direction of energy transfer in the dark sector: $Q>0$ corresponds to the decay of dark energy into dark matter, whereas $Q<0$ indicates energy flow from dark matter to dark energy.

We now analyze the evolution of matter density perturbations within the IDE framework. Throughout this work, we adopt the following assumptions:

\begin{itemize}
\item We restrict our analysis to the quasi-static and sub-horizon regimes, i.e., $k \gg aH$.
\item We neglect dark energy clustering effects and assume that dark energy behaves as a vacuum component at the background level, with equation-of-state parameter $w=-1$. As standard and usual procedure, we following the formulation presented in Ref.~\cite{LimaEtAl97}, we model dark matter as a perfect fluid characterized by an energy density $\rho_{\rm DM}$ and four-velocity $u^\mu_{\rm DM}$, evolving under the influence of the gravitational potential $\Phi$.
\end{itemize}

%We adopt the Neo-Newtonian formalism, which provides a description of the non-relativistic dark matter fluid interacting with dark energy, characterized by an equation of state parameter $w = -1$ \cite{Brito:2024bhh}. Following the formulation discussed in \cite{LimaEtAl97}, we treat dark matter as a perfect fluid with energy density $\rho_{\text{DM}}$, four-velocity $u_{\text{DM}}$, and subject to a gravitational potential $\Phi$. 

Under these conditions, the evolution of dark matter can be described by the standard set of fluid equations—namely, the continuity, Euler, and Poisson equations—which take the form

\begin{align}\label{epneo1}
   &\left( \dfrac{\partial \rho_{\text{DM}}}{\partial t}\right)_r+ \vec{\nabla}_r\cdot\big(\rho_{\text{DM}}\vec{u}_{\text{DM}}\big) = Q\,,\\\label{epneo2}
      &\left( \dfrac{\partial \vec{u}_{\text{DM}}}{\partial t}\right)_r+ (\vec{u}_{\text{DM}}\cdot\vec{\nabla}_r)\vec{u}_{\text{DM}} = -\vec{\nabla}_r\Phi\,,\\\label{epneo3}
        &{\nabla}_r^2\Phi = 4\pi G\rho_{\text{DM}} -\Lambda\,,
\end{align}
where $Q$ represents the phenomenological interaction term. 

We begin by performing a change of variables from proper coordinates $(\vec{r},t)$ to comoving coordinates $(\vec{x},t)$, defined by the relations:
\begin{align}
     &\vec{r} = a(t)\vec{x}\,,\\
    &\vec{\nabla}_r = \dfrac{1}{a}\vec{\nabla}_x \,,\\
     &\left( \dfrac{\partial }{\partial t}\right)_r = \left( \dfrac{\partial }{\partial t}\right)_x - H\left(\vec{x}\cdot\vec{\nabla}_x \right)\,.
\end{align}

We consider small perturbations around the homogeneous and isotropic background, which are defined as
\begin{align}
\vec{u}_{\rm DM}(\vec{x},t) &= \dot{a}\,\vec{x} + \vec{v}_{\rm DM}(\vec{x},t) \,, \\
\rho_{\rm DM}(\vec{x},t) &= \bar{\rho}_{\rm DM}(t) + \delta\rho_{\rm DM}(\vec{x},t)
= \bar{\rho}_{\rm DM}(t)\left(1+\delta\right) \,, \\
\Phi(\vec{x},t) &=
\phi(\vec{x},t)
+ \frac{2\pi G}{3}\,\bar{\rho}_{\rm DM}(t)\,a^2 x^2
- \frac{\Lambda}{6}\,a^2 x^2 \,,
\end{align}
where $\delta \equiv \delta\rho_{\rm DM}/\bar{\rho}_{\rm DM}$ denotes the dark matter density contrast, $\vec{v}_{\rm DM}$ is the peculiar velocity field, and $\phi$ represents the perturbed gravitational potential. Since dark matter is treated as a pressureless fluid, its effective sound speed vanishes.

Substituting the perturbed quantities into the system \eqref{epneo1}-\eqref{epneo3}, we obtain the equations governing the evolution of the perturbations:
\begin{align}
& \left( \dfrac{\partial \delta }{\partial t}\right) + \dfrac{1}{a}\vec\nabla\cdot\vec{v}_{\text{DM}} + \dfrac{1}{a}\vec\nabla\cdot(\delta\vec{v}_{\text{DM}}) = -\frac{Q}{\bar\rho_{\text{DM}}}\delta\,,\\
  &\left(\dfrac{\partial \vec{v}_{\text{DM}}}{\partial t}\right)+H\vec{v}_{\text{DM}}+\frac{1}{a}(\vec{v}_{\text{DM}}\cdot\vec{\nabla})\vec{v}_{\text{DM}} = -\frac{\vec{\nabla}\phi}{a}\,,\\
     &{\nabla}^2\phi = 4\pi G\bar{\rho}_{\text{DM}}a^2\delta \,.
\end{align}

By linearizing these equations and neglecting second-order perturbative terms, we obtain:

\begin{align}\label{dddtper}
& \dot{\delta} +\dfrac{\vec\nabla\cdot\vec{v}_{\text{DM}}}{a} = -\frac{Q}{\bar\rho_{\text{DM}}}\delta\,,\\\label{dvdtpe}
  &\dot{\vec{v}}_{\text{DM}}+H\vec{v}_{\text{DM}} = -\frac{\vec{\nabla}\phi}{a}\,,\\
     &{\nabla}^2\phi = 4\pi G\bar{\rho}_{\text{DM}}a^2\delta \,.\label{naper}
\end{align}

Taking the divergence of Eq. \eqref{dvdtpe} and combining it with the time derivative of Eq. \eqref{dddtper} and the Poisson equation \eqref{naper}, we derive the second-order differential equation for the density contrast:
 \begin{align}\label{delta2T}
 \ddot{\delta} + \left( 2H+ \frac{Q}{\bar\rho_{\text{DM}}}\right)\dot{\delta} + \left[2H\frac{Q}{\bar\rho_{\text{DM}}}+  \frac{\partial}{\partial t}\left(\frac{Q}{\bar\rho_{\text{DM}}}\right)-4\pi G\bar{\rho}_{\text{DM}}\right]\delta = 0\,.
 \end{align}
 
For the purpose of numerical integration, Eq.~\eqref{delta2T} can be recast in terms of the scale factor $a$ as the independent variable. Making use of the identity $\partial_t = aH\,\partial_a$, we arrive at
\begin{align}\label{eqdelta}
    \delta'' + \left[ \frac{3}{a} + \frac{H'}{H} + \frac{Q}{a H \bar{\rho}_{\text{DM}}} \right]\delta' + 
    \left[ \frac{Q'}{a H \bar{\rho}_{\text{DM}}} + \frac{Q}{\bar{\rho}_{\text{DM}} H^2}\left(\frac{H'}{a} + \frac{2H}{a^2} \right) - \frac{4\pi G \bar{\rho}_{\text{DM}}}{a^2 H^2} \right]\delta = 0\,,
\end{align}
where the prime denotes a derivative with respect to the scale factor $a$.

We investigate the evolution of density perturbations for two specific interaction forms introduced in the Introduction. The first, denoted as Model~I, is characterized by the interaction term $Q_{\text{I}}=\varepsilon\,aH\,\bar{\rho}_{\Lambda}$. The second, referred to as Model~II, is defined by $Q_{\text{II}}=\varepsilon\,H\,\bar{\rho}_{\Lambda}$. The analysis of Model~I is presented in the main text, while the results for Model~II are deferred to Appendix~\ref{APA}. For comparison, we also solve Eq.~\eqref{eqdelta} within the framework of the standard spatially flat $\Lambda$CDM model. It should be noticed that Models~I and~II differ solely in the specific phenomenological parameterization adopted for the interaction term $Q$, related by $Q_{\text{I}} = a Q_{\text{II}}$. From a physical perspective, both models describe the same underlying mechanism of energy exchange between dark energy and dark matter, leading to nearly identical cosmological dynamics. The distinction between them is essentially technical, reflecting slightly different ansatzes for the time dependence of the interaction rate. Notably, at the present epoch ($a=1$), the interaction terms coincide exactly. Furthermore, their evolution is intrinsically linked by the Hubble rate, according to the relation:\begin{equation}\frac{d}{dt}\ln\left(\frac{Q_{\text{I}}}{Q_{\text{II}}}\right) = H \,.\end{equation}As anticipated, the statistical results obtained for both models differ only by negligible statistical fluctuations. Consequently, these models should be regarded primarily as consistency tests of the IDE parameterization, rather than as physically distinct cosmological scenarios (more details in the Appendix \ref{APA}).

%Models~I and~II differ only in the specific phenomenological parameterization adopted for the interaction term $Q$. From a physical perspective, both models describe the same underlying scenario of energy exchange between dark energy and dark matter and therefore lead to very similar cosmological dynamics. The distinction between them is essentially technical, reflecting different but closely related choices for the time dependence of the interaction rate. As such, these models should be regarded primarily as basic consistency tests of the IDE parameterization, rather than as physically distinct cosmological $Q$ scenarios.
 
\subsection{Model~I: \texorpdfstring{$Q_{\text{I}}=\varepsilon\,aH\,\bar{\rho}_{\Lambda(t)}$}{Q = eps aH rho\_Lambda}}

We consider a spatially flat $\Lambda(t)$CDM scenario characterized by the phenomenological interaction term
\begin{equation}
Q_{\text{I}} = \varepsilon\,aH\,\bar{\rho}_{\Lambda(t)} \,.
\end{equation}
Although the continuity equations~\eqref{DMQ} and~\eqref{DEQ} are naturally written in terms of cosmic time, it is analytically more convenient to recast them using the scale factor $a$ as the independent variable. Making use of the relation $d/dt = aH\,d/da$,
the background evolution equations for the dark sector energy densities can be written as
\begin{align}\label{rhoMidea}
   a   \dfrac{\partial \bar\rho_{\text{DM}} }{\partial a}+3\bar\rho_{\text{DM}} &= \varepsilon a \bar\rho_{\Lambda(t)}\,,\\\label{rhoEidea}
     \dfrac{\partial \bar\rho_{\Lambda(t)} }{\partial a} &=-\varepsilon \bar\rho_{\Lambda(t)}\,.
\end{align}

The solution to the differential equation for dark energy \eqref{rhoEidea} is straightforward:
\begin{align}\label{eqrhoDEI}
    \bar\rho_{\Lambda(t)} = \bar\rho_{\Lambda 0}e^{\varepsilon(1-a)}\,.
\end{align}

Substituting this result into Eq. \eqref{rhoMidea}, the evolution equation for dark matter becomes:
\begin{align}
      a\dfrac{\partial \bar\rho_{\text{DM}} }{\partial a}+3\bar\rho_{\text{DM}} &= \varepsilon a \bar\rho_{\Lambda 0}e^{\varepsilon(1-a)}\,.
\end{align}

Solving this linear differential equation, we obtain the expression for the dark matter energy density:
\begin{align}\label{eqrhoDMI}
    \bar\rho_{\text{DM}}(a)=\frac{\bar\rho_{\Lambda0}}{a^3\varepsilon^3}\left[-e^{\varepsilon(1-a)} A+B\right]+\frac{\bar\rho_{\text{DM}0}}{a^3}\,,
\end{align}
where, for compactness, we define the polynomial structures $A$ and $B$ as:
\begin{align}
    A &\equiv a^3 \varepsilon^3 + 3 a^2 \varepsilon^2 + 6 a \varepsilon + 6\,,\\
    B &\equiv \varepsilon^3 + 3 \varepsilon^2 + 6 \varepsilon + 6\,.
\end{align}

By substituting the solutions \eqref{eqrhoDMI} and \eqref{eqrhoDEI} into the Friedmann equation, the dimensionless Hubble rate, $E(a) \equiv H(a)/H_0$, takes the form:
\begin{align}
    E(a)^2 = \frac{e^{-a\varepsilon}}{a^4 \varepsilon^3} \left[
    e^{a\varepsilon} \left( a\varepsilon^3 \Omega_{M0} + a \Omega_{\Lambda 0} B + \varepsilon^3 \Omega_{r0} \right)
    - 3 a \Omega_{\Lambda 0} e^{\varepsilon} C
    \right]\,,
\end{align}
which depends on the polynomial term $C$, defined as:
\begin{align}
    C &\equiv a^2 \varepsilon^2 + 2 a \varepsilon + 2\,.
\end{align}

For the numerical analysis, we define the total matter density parameter as
\begin{equation}
\Omega_{M0} \equiv \Omega_{\mathrm{DM}0} + \Omega_{b0}\,.
\end{equation}
Since the observational data sets employed in this work are predominantly geometric and restricted to low redshifts, they primarily constrain the total matter content rather than its individual components.

Turning to the linear perturbation analysis, and adopting the interaction term $Q_{\text{I}} = \varepsilon\,aH\,\bar{\rho}_{\Lambda}$,
the second-order differential equation governing the evolution of the matter density contrast Eq.~\eqref{eqdelta}, can be written explicitly as

\begin{equation}\label{eqdeltan1}
    \begin{split}
        \delta'' = & -\delta' \left[
        \frac{3}{a} - \frac{a^3 \varepsilon^4 \Omega_{\Lambda 0} e^{\varepsilon}}{\mathcal{D}_1} + \frac{1}{2a}\frac{\mathcal{D}_3}{\mathcal{D}_2}
        \right]  - \delta \frac{\varepsilon^3}{2a} \left[
         -\frac{4 a^3 \varepsilon \Omega_{\Lambda 0} e^{\varepsilon}}{\mathcal{D}_1}
         + \frac{3 \mathcal{D}_1}{\varepsilon^3 \mathcal{D}_2}
         + \frac{a^3 \varepsilon \Omega_{\Lambda 0} e^{\varepsilon}}{\mathcal{D}_1^2} \mathcal{K}
        \right]\,,
    \end{split}
\end{equation}
 
The auxiliary function $\mathcal{K}$ is defined as:
\begin{equation}
    \mathcal{K} = 6 a \varepsilon \Omega_{\Lambda 0} e^{\varepsilon} C - 8 \mathcal{D}_1 - 2 a \varepsilon e^{a\varepsilon}\left(B \Omega_{\Lambda 0} + \varepsilon^3 \Omega_{DM0}\right) - \frac{\mathcal{D}_1 \mathcal{D}_3}{\mathcal{D}_2}\,.
\end{equation}

To simplify the equations, we have introduced the functions $\mathcal{D}_i$, utilizing the polynomials $A, B,$ and $C$ defined previously:
\begin{align}
    \mathcal{D}_1 &\equiv e^{\varepsilon} \Omega_{\Lambda 0} A - e^{a \varepsilon} \left(B \Omega_{\Lambda 0} + \varepsilon^3 \Omega_{DM0}\right)\,, \\
    \mathcal{D}_2 &\equiv e^{a \varepsilon} \left[ a \left(B \Omega_{\Lambda 0} + \varepsilon^3 \Omega_{M0}\right) + \varepsilon^3 \Omega_{r0} \right] - 3 a e^{\varepsilon} \Omega_{\Lambda 0} C\,, \\
    \mathcal{D}_3 &\equiv 3 a e^{\varepsilon} \Omega_{\Lambda 0} A - e^{a\varepsilon} \left[ 3 a \left(B \Omega_{\Lambda 0} + \varepsilon^3 \Omega_{M0}\right) + 4 \varepsilon^3 \Omega_{r0} \right]\,.
\end{align}

\section{Observational Data and Statistical Methodology}
\label{data}

Observational data play a central role in testing cosmological models and assessing the significance of current tensions. The high precision achieved by modern data sets enables stringent constraints on model parameters, allowing us to identify which $\Lambda(t)$CDM scenarios provide the most accurate description of the observed Universe. Below, we describe the observational data samples employed in this analysis:

\begin{itemize}
\item \textbf{Redshift-Space Distortions (RSD):}
Redshift-space distortion measurements originate from the peculiar velocities of galaxies, which induce anisotropies in the observed clustering pattern when distances are inferred from redshifts. As a result, RSD provides a direct probe of the growth of cosmological structures across cosmic time and on large, quasi-linear scales \cite{Kaiser:1987qv}.

A key quantity used to constrain cosmological models is the linear growth rate,
\begin{equation}
f(a) \equiv \frac{d\ln \delta(a)}{d\ln a},
\end{equation}
where $\delta \equiv \delta\rho_{\rm DM}/\bar{\rho}_{\rm DM}$ denotes the dark matter density contrast and $\bar{\rho}_{\rm DM}$ is the background dark matter density. In practice, RSD observations constrain the combination $f\sigma_8(z)$, where $\sigma_8$ represents the root-mean-square amplitude of matter fluctuations smoothed over spheres of radius $8\,h^{-1}\,\mathrm{Mpc}$. This observable is particularly robust, as it is largely insensitive to galaxy bias and avoids the need for an independent determination of $\sigma_8$.

In this work, we employ a compilation of 20 independent RSD measurements of $f\sigma_8(z)$, assembled by \cite{Avila:2022xad}, covering the redshift range $0.02 < z < 1.944$. These data provide strong constraints on the growth history of matter perturbations.

\item \textbf{Type Ia Supernovae (SNe Ia):}
Type Ia Supernovae provide one of the most robust and direct probes of the late-time expansion history of the Universe through measurements of the luminosity distance--redshift relation. In this work, we consider three independent SNe Ia compilations, which are analyzed separately to avoid statistical redundancy due to shared observational data.

First, we use the Pantheon+ (PP) compilation \cite{Brout:2022vxf}, which comprises 1701 light curves from 1550 spectroscopically confirmed SNe Ia, covering the redshift range $0.001 < z < 2.26$. Second, we employ the Dark Energy Survey Year 5 sample (DESY5) \cite{DES:2024jxu}, which includes 1829 photometrically classified SNe Ia discovered over the full five-year DES program, spanning the interval $0.10 < z < 1.13$. Finally, we consider the Union3 compilation \cite{Rubin:2023jdq}, consisting of 2087 cosmologically useful SNe Ia assembled from 24 distinct datasets and covering the redshift range $0.001 < z < 2.26$.

Given the partial overlap among these datasets, each SNe Ia compilation is treated independently in our analysis to ensure statistical consistency.

\item \textbf{Baryon Acoustic Oscillations (BAO):} We employ the most recent Baryon Acoustic Oscillation (BAO) measurements from the DESI collaboration Data Release 2 (DR2) \cite{DESI:2025zgx}, which span the redshift interval $0.295 < z < 2.33$. This dataset provides measurements of the dimensionless ratios $D_{M}/r_{s,\mathrm{drag}}$, $D_{H}/r_{s,\mathrm{drag}}$, and $D_{V}/r_{s,\mathrm{drag}}$\footnote{Here, $D_{M}$ denotes the transverse comoving distance, $D_{H} \equiv c/H(z)$ is the Hubble distance, $D_{V}$ is the volume-averaged (isotropic) BAO distance, and $r_{s,\mathrm{drag}}$ is the comoving sound horizon at the drag epoch. The drag-epoch redshift $z_{\mathrm{drag}}$ is computed using the approximation described in \cite{Hu:1995en}.}.

\item \textbf{Cosmic Chronometers (CCs):} The Hubble expansion rate, $H(z)$, can be directly inferred from the differential aging of passively evolving galaxies through the relation
\begin{equation}
H(z) = -\frac{1}{1+z}\,\frac{dz}{dt}\,.
\end{equation}

In this work, we employ the most recent compilation of Cosmic Chronometer (CC) data, which consists of 32 measurements of $H(z)$ with an associated covariance matrix, spanning the redshift range $0.07 < z < 1.97$ \cite{Moresco:2022phi}. These measurements are obtained by estimating the differential ages of galaxies selected to minimize contamination from recent star formation.

A key improvement of this dataset is the inclusion of a comprehensive assessment of systematic uncertainties. As detailed in Ref.~\cite{Moresco:2022phi}, the total error budget accounts for several sources of systematics, including uncertainties in metallicity, Star Formation History, assumptions in Stellar Population Synthesis models, and the so-called rejuvenation effect. This makes the CC data set particularly robust for constraining the late-time expansion history of the Universe.

%\item \textbf{CMB Distance Priors (PCMB):} We adopt the Cosmic Microwave Background (CMB) distance priors derived from the Planck 2018 data release \cite{Planck:2018dkp}. As shown in Ref.~\cite{Chen:2018dbv}, CMB distance priors encapsulate the dominant geometrical information of the CMB, providing robust constraints on background cosmological parameters while avoiding the computational cost of solving the full Boltzmann hierarchy for the temperature and polarization power spectra.

%In particular, we employ the priors on the physical baryon density $\Omega_{b0}h^2$, the shift parameter $R$, and the acoustic angular scale $\ell_A$, defined following the formalism of Ref.~\cite{Chen:2018dbv}. These quantities efficiently constrain the comoving distance to the last scattering surface and the expansion history up to recombination, making the PCMB dataset especially suitable for joint analyses with low-redshift geometric probes.

\textbf{CMB Distance Priors (PCMB):} We adopt the Cosmic Microwave Background (CMB) distance priors derived from the Planck 2018 data release \cite{Planck2018}. As shown in Ref.~\cite{Chen:2018dbv}, CMB distance priors encapsulate the dominant geometrical information of the CMB, providing robust constraints on background cosmological parameters while avoiding the computational cost of solving the full Boltzmann hierarchy for the temperature and polarization power spectra. In particular, we employ the priors on the physical baryon density $\Omega_{b0}h^2$, the shift parameter $R$, and the acoustic angular scale $\ell_A$, defined in.~\cite{Chen:2018dbv}. These quantities efficiently constrain the comoving distance to the last scattering surface and the expansion history up to recombination, making the PCMB dataset especially suitable for joint analyses with low-redshift geometric probes.

It is important to emphasize that the CMB distance priors are extracted under the assumption of the standard $\Lambda$CDM cosmological model. Consequently, when applied to extended scenarios—such as IDE models—the use of PCMB introduces an implicit model dependence. In particular, these priors may not fully capture the impact of non-standard physics on the evolution of perturbations or on the detailed shape of the CMB power spectra. For this reason, in the context of IDE models, the PCMB dataset should be interpreted mainly as an effective means to alleviate statistical degeneracies among background parameters—such as $\Omega_{m0}$, $H_0$, and dark energy parameters—rather than as a fully model constraint, which is the specific purpose for which these estimates are employed in this work.

\end{itemize}

The models considered in this work are analyzed within a Bayesian statistical framework using Markov Chain Monte Carlo (MCMC) techniques \cite{Gregory2005,HobsonEtAl14,GelmanEtAl14,GilksRich96}. We assume a Gaussian likelihood of the form $\mathcal{L} \propto \exp(-\chi^2/2)$ \cite{PressEtAl1992}, where the $\chi^2$ function can be written in matrix notation as:

\begin{equation}
    \chi^2 = (\mathbf{Y} - \mathbf{Y}_{\text{obs}}) \mathbf{C}^{-1} (\mathbf{Y} - \mathbf{Y}_{\text{obs}})^T \,,
\end{equation}
where $\mathbf{Y}$ represents the theoretical prediction vector, $\mathbf{Y}_{\text{obs}}$ is the observation vector, and $\mathbf{C}$ is the covariance matrix of the data sample. According to Bayes' Theorem, the posterior distribution $p(\theta_j)$ for the free parameters $\theta_j$ is given by:
\begin{equation}
    p(\theta_j) \propto \pi(\theta_j) \mathcal{L}(x_i, \theta_j) \,,
\end{equation}
where $\pi(\theta_j)$ is the prior distribution and $\mathcal{L}$ is the likelihood. We sample the posterior distribution to find the best-fit parameters and confidence intervals for each data sample, both separately and combined.

To perform the sampling, we use the publicly available software \texttt{emcee} \cite{ForemanEtAl13}, which implements the Affine Invariant Ensemble Sampler method \cite{GoodmanWeare10}. We assume flat priors for all free parameters in the $n$-dimensional parameter space.
The convergence of the MCMC chains was tested using the integrated autocorrelation time ($\tau$) provided by \texttt{emcee}. Following the documentation criteria\footnote{\url{https://emcee.readthedocs.io/en/stable/}}, we consider the chains to be converged when the number of steps satisfies $n_{\mathrm{step}} \gg \tau$. Specifically, we ensure $n_\mathrm{step} > 50\tau$ for all free parameters in the analyzed models.
Finally, the posterior distributions and confidence contours were plotted using the \texttt{getdist} software\footnote{\texttt{getdist} is part of the CosmoMC package \cite{cosmomc}.}.

%\begin{table}[H]
%\centering
%\begin{tabular} { c| c}
%\hline \hline
%Parameter & Range of the flat prior\\
%\hline
%{\boldmath$H_0$} (km/s/Mpc) & $[20,120]$\\
%{\boldmath$\Omega_{DM0} $} & $[0,1.0]$\\
%{\boldmath$\varepsilon $} & $[-1,1]$\\
%{\boldmath$\sigma_{80} $} & $[0.1,2.5]$\\
%\hline
%\end{tabular}
%\caption{Priors on the free parameters.}
%\label{PriorPPN1}
%\end{table}

\section{Results}
\label{Res}

We constrain the IDE models by estimating their free parameters through the Bayesian statistical framework described in Section~\eqref{data}. Our analysis combines multiple low- and intermediate-redshift observational probes as defined previously. We are primarily interested in RSD measurements, as they provide direct constraints on the growth of cosmic structures. To further break degeneracies among background cosmological parameters, we also incorporate BAO data and PCMB. Since the Type Ia supernova catalogs share overlapping datasets, each compilation is analyzed independently to avoid statistical redundancy. 

We adopt flat priors for all free parameters. The Hubble constant is allowed to vary within the range $H_0 \in  [20,120]~\mathrm{km\,s^{-1}\,Mpc^{-1}}$. The present-day dark matter density parameter is restricted to $\Omega_{\mathrm{DM}0} \in [0,1.0]$, while the interaction parameter spans the interval $\varepsilon \in [-1,1]$. Finally, the amplitude of matter fluctuations is constrained to $\sigma_8 \in [0.1,2.5]$.
\\

%\RCN{The results of our combined analysis are summarized in Table~\ref{tab:param_model1_lcdm}, showing constraints at the 68\% and 95\% confidence levels. In this section, we focus on the observational constraints for IDE Model I and its comparison with the standard flat $\Lambda$CDM scenario.}

%\textbf{Por favor, definir no texto em detalhes a nomenclatura usada: Early, Local, All. Ou seja, escrever o que significa cada coisa na tabela II.}
%\\
%\textbf{Essa nomenclatura não sera mais utilizada.}
% --- TABELA DE PARÂMETROS (MODELO I e LCDM) ---

\begin{table}[!htbp]
\centering
\footnotesize
\setlength{\tabcolsep}{4pt}
\renewcommand{\arraystretch}{1.05}

\begin{tabular}{lccccc}
\hline\hline
\textbf{Model/Dataset} & $H_0$ & $\Omega_{\mathrm{DM}0}$ & $\Omega_{b0}$ & $\varepsilon$ & $S_8$ \\
\hline
\multicolumn{6}{l}{\textbf{Model I}}\\
\hline
\quad PCMB+BAO & $68.67\pm0.30$ & $0.2524\pm0.0035$ & $0.04741\pm0.00040$ & $0.025\pm0.010$ & --- \\
\hline
\quad RSD+CCs+PP & $67.0^{+3.1}_{-3.5}$ & $0.280\pm0.031$ & $0.054^{+0.019}_{-0.034}$ & $0.0096^{+0.024}_{-0.030}$ & $0.896^{+0.044}_{-0.051}$ \\
\quad RSD+CCs+BAO+PP & $68.7\pm3.3$ & $0.260\pm0.011$ & $0.0464^{+0.0047}_{-0.0035}$ & $0.006\pm0.024$ & $0.863\pm0.031$ \\
%\quad PCMB+CCs+BAO+PP & $68.54\pm0.30$ & $0.2539\pm0.0035$ & $0.04756\pm0.00040$ & $0.024\pm0.010$ & $0.90\pm0.17$ \\
\quad RSD+CCs+PCMB+BAO+PP & $68.51\pm0.29$ & $0.2538\pm0.0035$ & $0.04768\pm0.00038$ & $0.0190\pm0.0093$ & $0.870\pm0.026$ \\
\hline
\quad RSD+CCs+UNION3 & $65.4\pm3.4$ & $0.306\pm0.037$ & $0.055^{+0.019}_{-0.034}$ & $0.024^{+0.028}_{-0.040}$ & $0.916\pm0.048$ \\
\quad RSD+CCs+BAO+UNION3 & $68.6\pm3.3$ & $0.260\pm0.011$ & $0.0463^{+0.0047}_{-0.0035}$ & $0.006\pm0.024$ & $0.863\pm0.031$ \\
%\quad PCMB+CCs+BAO+UNION3 & $68.56\pm0.30$ & $0.2536\pm0.0035$ & $0.04753\pm0.00040$ & $0.024\pm0.010$ & $0.90\pm0.17$ \\
\quad RSD+CCs+PCMB+BAO+UNION3 & $68.51\pm0.30$ & $0.2538\pm0.0035$ & $0.04767\pm0.00038$ & $0.0190\pm0.0091$ & $0.869\pm0.026$ \\
\hline
\quad RSD+CCs+DESY5 & $65.8\pm3.3$ & $0.300\pm0.030$ & $0.054^{+0.019}_{-0.034}$ & $0.020^{+0.028}_{-0.034}$ & $0.911^{+0.043}_{-0.050}$ \\
\quad RSD+CCs+BAO+DESY5 & $68.2\pm3.3$ & $0.268\pm0.011$ & $0.0457^{+0.0047}_{-0.0035}$ & $0.014\pm0.025$ & $0.870\pm0.031$ \\
%\quad PCMB+CCs+BAO+DESY5 & $68.42\pm0.29$ & $0.2552\pm0.0035$ & $0.04769\pm0.00040$ & $0.023\pm0.010$ & $0.91\pm0.17$ \\
\quad RSD+CCs+PCMB+BAO+DESY5 & $68.39\pm0.29$ & $0.2552\pm0.0035$ & $0.04780\pm0.00038$ & $0.0185\pm0.0093$ & $0.870\pm0.026$ \\
\hline
\multicolumn{6}{l}{\textbf{$\Lambda$CDM}} \\
\hline
\quad PCMB+BAO & $68.59\pm0.30$ & $0.2514\pm0.0035$ & $0.04794\pm0.00034$ & --- & --- \\
\hline
\quad RSD+CCs+PP & $67.1\pm3.2$ & $0.272\pm0.029$ & $0.058^{+0.032}_{-0.027}$ & --- & $0.895\pm0.046$ \\
\quad RSD+CCs+BAO+PP & $68.8\pm3.2$ & $0.2581\pm0.0094$ & $0.0465^{+0.0045}_{-0.0034}$ & --- & $0.859\pm0.026$ \\
%\quad PCMB+CCs+BAO+PP & $68.47\pm0.30$ & $0.2528\pm0.0035$ & $0.04806\pm0.00033$ & --- & $0.90\pm0.17$ \\
\quad RSD+CCs+PCMB+BAO+PP & $68.47\pm0.29$ & $0.2528\pm0.0034$ & $0.04807\pm0.00033$ & --- & $0.859\pm0.026$ \\
\hline
\quad RSD+CCs+UNION3 & $66.0\pm3.4$ & $0.289\pm0.032$ & $0.060^{+0.033}_{-0.020}$ & --- & $0.907\pm0.046$ \\
\quad RSD+CCs+BAO+UNION3 & $68.8\pm3.3$ & $0.258\pm0.010$ & $0.0465^{+0.0046}_{-0.0034}$ & --- & $0.859\pm0.026$ \\
%\quad PCMB+CCs+BAO+UNION3 & $68.48\pm0.30$ & $0.2526\pm0.0035$ & $0.04805\pm0.00034$ & --- & $0.90\pm0.17$ \\
\quad RSD+CCs+PCMB+BAO+UNION3 & $68.48\pm0.30$ & $0.2526\pm0.0035$ & $0.04805\pm0.00033$ & --- & $0.859\pm0.025$ \\
\hline
\quad RSD+CCs+DESY5 & $66.0\pm3.3$ & $0.289\pm0.028$ & $0.060^{+0.033}_{-0.020}$ & --- & $0.907\pm0.044$ \\
\quad RSD+CCs+BAO+DESY5 & $68.4\pm3.3$ & $0.2648\pm0.0096$ & $0.0458^{+0.0047}_{-0.0034}$ & --- & $0.862\pm0.026$ \\
%\quad PCMB+CCs+BAO+DESY5 & $68.36\pm0.29$ & $0.2541\pm0.0035$ & $0.04818\pm0.00033$ & --- & $0.90\pm0.17$ \\
\quad RSD+CCs+PCMB+BAO+DESY5 & $68.36\pm0.29$ & $0.2541\pm0.0034$ & $0.04817\pm0.00033$ & --- & $0.861\pm0.025$ \\
\hline\hline
\end{tabular}

\caption{68\% confidence level constraints on the free parameters of Model~I and the $\Lambda$CDM model from different cosmological dataset combinations.}
\label{tab:param_model1_lcdm}
\end{table}

We present here the constraints on Model~I of IDE, characterized by the interaction parameter $\varepsilon$, alongside the standard $\Lambda$CDM model. Table~\ref{tab:param_model1_lcdm} summarizes the inferred values for the Hubble constant $H_0$, the present-day dark matter density $\Omega_{\rm DM0}$, the baryon density $\Omega_{b0}$, the interaction parameter $\varepsilon$, and the clustering parameter $S_8$ at 68\% confidence level.

For the IDE model, the interaction parameter $\varepsilon$ is consistently constrained to be small across all dataset combinations, indicating a mild preference for a weak interaction. High-redshift datasets, such as PCMB+BAO, provide $\varepsilon = 0.025 \pm 0.010$, while including low-redshift probes such as RSD, CCs, and SNe Ia leads to slightly weaker constraints, e.g., $\varepsilon = 0.0190 \pm 0.0093$ for the full combination RSD+CCs+PCMB+BAO+Pantheon+. This demonstrates that combining background expansion data with growth-of-structure measurements improves the precision on the interaction parameter.

The inferred Hubble constant $H_0$ in the IDE scenario ranges from $\sim 65$ km/s/Mpc to $\sim 68.7$ km/s/Mpc depending on the dataset. Combinations including CMB priors yield $H_0 \sim 68.5$ km/s/Mpc, consistent with Planck results under $\Lambda$CDM, whereas low-redshift datasets alone tend to produce slightly lower values. The present-day dark matter density $\Omega_{\rm DM0}$ shows mild variation, with high-redshift datasets favoring $\Omega_{\rm DM0} \sim 0.253$, while analyses using only low-redshift probes allow higher values up to $\Omega_{\rm DM0} \sim 0.306$. The baryon density $\Omega_{b0}$ remains well constrained, particularly when including CMB priors, around $\Omega_{b0} \sim 0.0476$, reflecting the robustness of baryon abundance determinations against the introduction of a dark sector interaction.

Regarding structure growth, the clustering parameter $S_8$ spans $0.86 \lesssim S_8 \lesssim 0.92$ in the IDE model, depending on the dataset combination. The inclusion of RSD measurements, which directly probe the growth of cosmic structures, significantly reduces uncertainties in $S_8$ compared to using only background data. 

Comparing with $\Lambda$CDM, the inferred values of $H_0$ and $\Omega_{\rm DM0}$ are similar when CMB data are included. As expected, $\varepsilon$ is absent in the standard model, and the derived $S_8$ values are largely consistent with those in the IDE scenario, although slightly lower for some low-redshift dataset combinations. This comparison highlights that a small dark sector interaction does not substantially alter background or growth observables but provides additional flexibility that may help alleviate potential tensions between high- and low-redshift measurements.

Finally, the choice of SNe Ia compilation (Pantheon+, Union3, DESY5) has only a minor impact on the inferred cosmological parameters, with differences well within 1$\sigma$ uncertainties. This demonstrates the robustness of the constraints and the consistency among the various supernova datasets when combined with RSD, CCs, and BAO measurements. Overall, the IDE model with a small positive interaction parameter $\varepsilon$ provides an excellent fit to the combined datasets, yielding results broadly consistent with $\Lambda$CDM while offering a minimal extension to the standard cosmological scenario.

\begin{figure}
    \centering
    \includegraphics[width=0.5\linewidth]{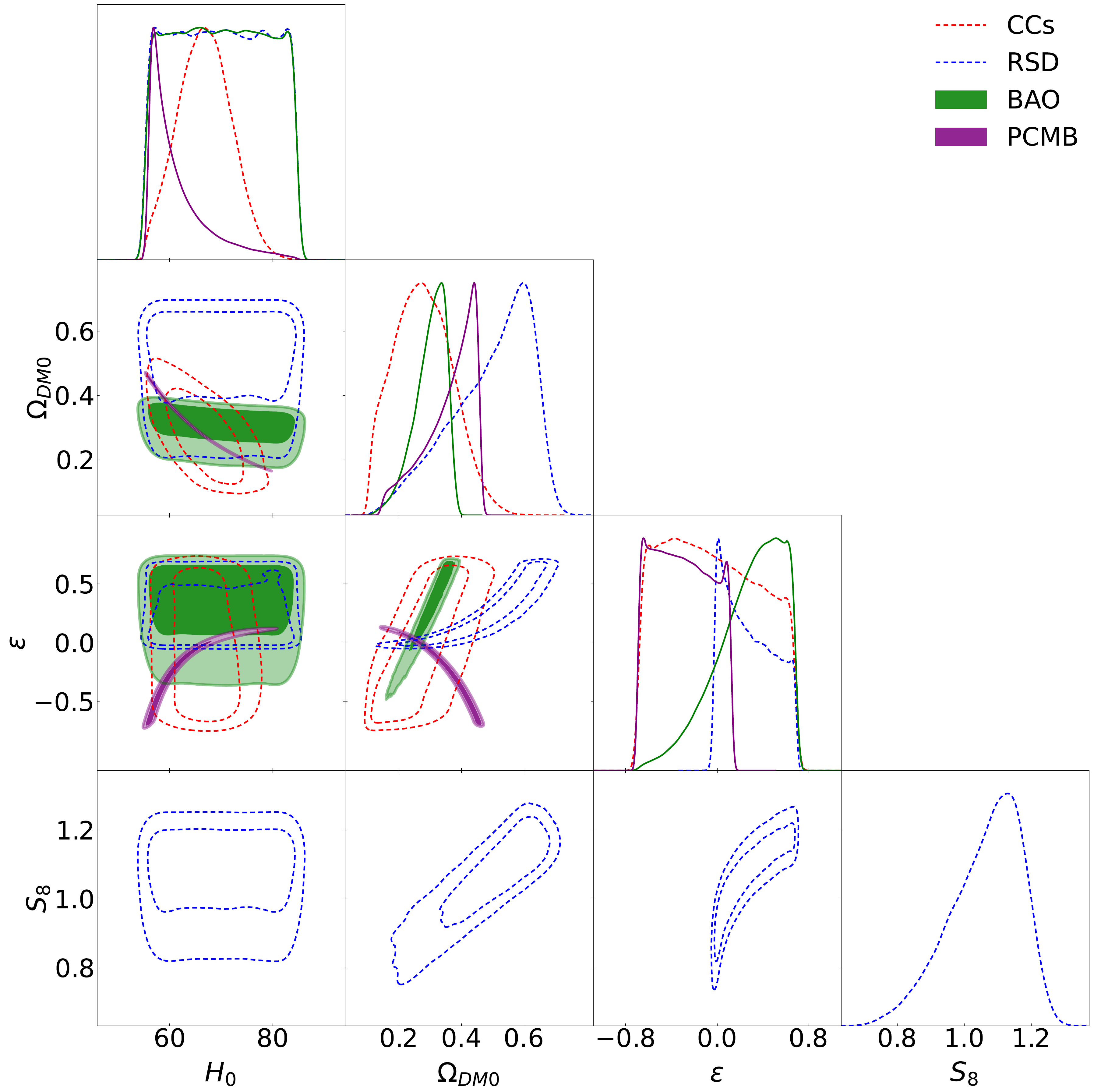}\includegraphics[width=0.5\linewidth]{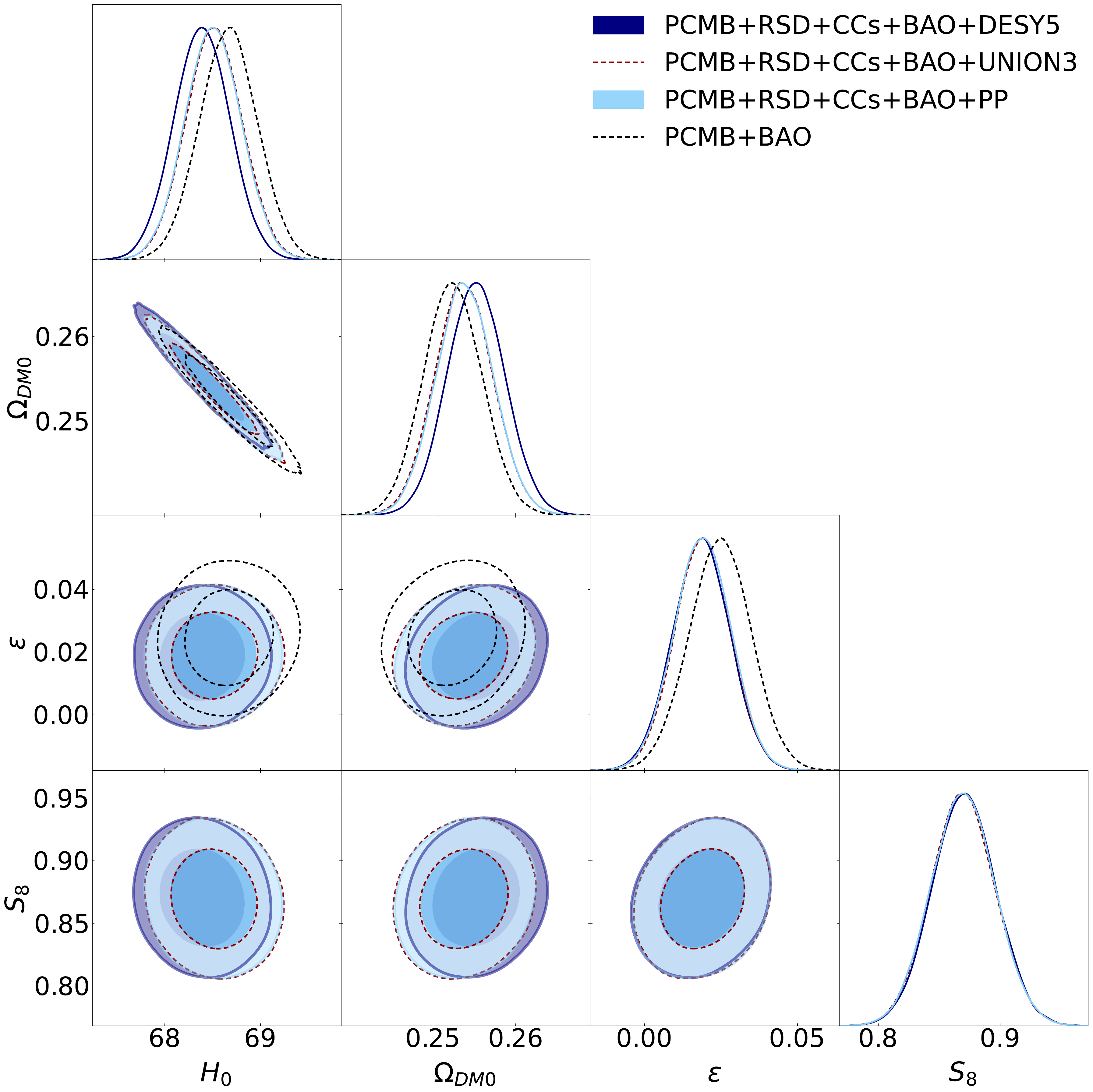}
    \caption{Marginalized posterior distributions and 68\% and 95\% confidence level (CL) contours for the main cosmological parameters $H_0$, $\Omega_{\rm DM0}$, $\varepsilon$, and $S_8$. The results are shown for different combinations of datasets, as indicated in the legend. \textbf{Left panel:} Constraints from individual datasets. \textbf{Right panel:} Constraints from the joint analysis of datasets.}
    \label{fig:contoursDESY5_N1}
\end{figure}

A quantitative comparison between the IDE model and $\Lambda$CDM reveals subtle but noteworthy differences in key cosmological parameters. Considering combinations that include high-redshift CMB priors, the inferred Hubble constant in IDE is $H_0 = 68.51 \pm 0.29$ km/s/Mpc, slightly higher than the corresponding $\Lambda$CDM value of $H_0 = 68.47 \pm 0.29$ km/s/Mpc, although both are consistent within $1\sigma$. For low-redshift datasets alone, such as RSD+CCs+Pantheon+, IDE yields $H_0 = 67.0^{+3.1}_{-3.5}$ km/s/Mpc compared to $\Lambda$CDM's $H_0 = 67.1 \pm 3.2$ km/s/Mpc, showing negligible differences.  

Similarly, the clustering parameter $S_8$ is mildly affected by the interaction. For the full dataset combination RSD+CCs+PCMB+BAO+Pantheon+, IDE produces $S_8 = 0.870 \pm 0.026$, slightly higher than the $\Lambda$CDM value $S_8 = 0.859 \pm 0.026$. Low-redshift datasets alone tend to give slightly higher $S_8$ in IDE as well, e.g., $S_8 = 0.896^{+0.044}_{-0.051}$ versus $\Lambda$CDM's $S_8 = 0.895 \pm 0.046$ for RSD+CCs+Pantheon+. These results indicate that while the IDE extension introduces only small shifts in $H_0$ and $S_8$, the model slightly favors higher values of both parameters.

Figure \ref{fig:contoursDESY5_N1} shows the marginalized $1\sigma$ and $2\sigma$ confidence contours along with the posterior distributions for the free parameters of Model~I. The left panel displays constraints from individual datasets, including CCs, RSD, BAO, and PCMB, while the right panel shows the combined constraints for various multi-probe combinations. The results indicate a high level of consistency among the different datasets, as the posterior distributions for $H_0$, $\Omega_{\rm DM0}$, $\Omega_{b0}$, $\varepsilon$, and $S_8$ largely overlap. As more datasets are combined, the contours become progressively tighter, reflecting a significant improvement in the precision of the parameter estimates, particularly for $H_0$ and $\Omega_{\rm DM0}$. Mild negative correlations are observed between $H_0$ and $\Omega_{\rm DM0}$, in line with geometrical degeneracies, while the interaction parameter $\varepsilon$ exhibits weak correlations with both $H_0$ and $\Omega_{\rm DM0}$, indicating that its effects on the expansion history are subtle. The clustering parameter $S_8$ appears largely uncorrelated with $\Omega_{b0}$ and only mildly correlated with $\varepsilon$ and $\Omega_{\rm DM0}$, consistent with the fact that growth-of-structure probes primarily constrain $S_8$ independently of baryon abundance. In short, Model~I is consistent with current cosmological observations, and that the combination of low- and high-redshift datasets provides precise and internally consistent constraints on both the standard baseline parameters and the dark sector interaction parameter $\varepsilon$.

In Ref.~\cite{Bernui:2023byc}, the authors investigated an IDE scenario featuring an interaction term of the form $Q=\varepsilon a H \rho_{\rm DE}$, which directly corresponds to our Model~I when identifying $\rho_{\rm DE}=\bar{\rho}_{\Lambda(t)}$. Using a combination of CMB data and transverse BAO measurements, they obtained the constraints $\varepsilon>-0.207$, $H_0 = 68.93^{+0.79}_{-1.20}\,\mathrm{km\,s^{-1}\,Mpc^{-1}}$, and $S_8 = 0.891^{+0.025}_{-0.062}$. 
Our results for Model~I, derived from the joint analysis of RSD, CCs, PCMB, BAO, and PP data, show remarkable agreement with these findings. In particular, our estimate $S_8 = 0.870 \pm 0.026$ is fully consistent with their constraint at the ${\sim}1\sigma$ level, indicating that IDE models with interactions proportional to $aH\rho_{\Lambda}$ naturally accommodate relatively high clustering amplitudes when confronted with large-scale structure data.

More recently, Ref.~\cite{Sabogal:2024yha} explored IDE models incorporating RSD information, explicitly distinguishing between scenarios in which energy is transferred from dark matter to dark energy ($\xi<0$) and the opposite case ($\xi>0$). Using the data combination DESI DR1 BAO~\cite{DESI:2024mwx}, PP, RSD, and CCs, they reported significantly lower clustering amplitudes than those obtained in our analysis, with $S_8 = 0.775 \pm 0.026$ for the negative coupling case and $S_8 = 0.812^{+0.028}_{-0.032}$ for the positive coupling case. 
Furthermore, when CMB data were included in their full analysis for the $\xi>0$ scenario, an even stronger suppression of structure growth was found, yielding $S_8 = 0.798 \pm 0.010$. In comparison, our result $S_8 \simeq 0.870$ lies systematically above these values. Quantitatively, we find a tension at the level of ${\sim}1.5\sigma$ relative to their non-CMB result for $\xi>0$, which increases to ${\sim}2.6\sigma$ when compared with their full analysis including CMB data. This difference highlights the sensitivity of IDE constraints to both the assumed interaction scheme and the treatment of perturbations.

Finally, Ref.~\cite{Silva:2025hxw} analyzed a traditional IDE framework using BAO measurements combined with Planck CMB data and several Type Ia supernova samples. They consistently found lower clustering amplitudes, reporting $S_8 = 0.8242^{+0.0095}_{-0.018}$ for the combination with PP, $S_8 = 0.827^{+0.011}_{-0.018}$ with Union3, and $S_8 = 0.8165^{+0.0084}_{-0.011}$ with DESY5. In contrast, our analysis incorporates RSD and CCs in addition to PCMB, BAO, and SNe~Ia, and systematically favors higher values of the clustering parameter. Across all supernova samples considered, we obtain $S_8 \simeq 0.870 \pm 0.026$. 
The comparison reveals a mild but persistent tension between our results and those of Ref.~\cite{Silva:2025hxw}, ranging from $1.4\sigma$ for Union3 and $1.6\sigma$ for Pantheon+ to $1.9\sigma$ for DESY5.

It is worth emphasizing that the three reference works discussed above~\cite{Bernui:2023byc,Sabogal:2024yha,Silva:2025hxw} employ a fully relativistic treatment of cosmological perturbations and make use of the complete full CMB likelihood information. By contrast, our analysis is based on a Newtonian description of perturbations and relies on CMB geometric distance priors rather than the full CMB data. These methodological differences may partially account for the systematic shift toward higher $S_8$ values found in our study, underscoring the importance of the perturbative framework and data choice in assessing the impact of IDE scenarios on the growth of cosmic structures.

\section{\label{Con}Conclusions}

This work was motivated by the persistent tensions within the $\Lambda$CDM paradigm, particularly the $S_8$ tension in light of the RSD measurements, which highlights a discrepancy between early- and late-Universe measurements of large scale structure formation. We explored a minimal extension of the standard model by considering an IDE scenario within a spatially flat $\Lambda(t)$CDM cosmology. Specifically, we investigated phenomenological interaction terms of the form $Q_{\text{I}} = \epsilon a H \bar{\rho}_{\Lambda}$ (Model I) and $Q_{\text{II}} = \epsilon H \bar{\rho}_{\Lambda}$ (Model II), which describe an energy transfer between a vacuum dark energy component and dark matter.

Our analysis was conducted at both the background and linear perturbation levels (quasi-static approximation and sub-horizon modes). To constrain the models observationally, we performed a robust Bayesian statistical analysis using a comprehensive combination of current cosmological probes. The main results for Model I, using the full dataset combination (RSD+CCs+PCMB+BAO+Pantheon+), yielded:
\begin{itemize}
    \item A Hubble constant of $H_0 = 68.51 \pm 0.29 \ \mathrm{km \ s^{-1} \ Mpc^{-1}}$,
    \item A dark matter density parameter of $\Omega_{\mathrm{DM0}} = 0.2538 \pm 0.0035$,
    \item A small but non-zero interaction parameter $\epsilon = 0.0190 \pm 0.0093$,
    \item A clustering parameter $S_8 = 0.870 \pm 0.026$.
\end{itemize}

For Model II, the constraints were statistically indistinguishable, yielding $S_8 = 0.872\pm 0.026$, which underscores the robustness of the perturbative dynamics with respect to the specific form of the interaction kernel.

The derived value of $S_8$ in both IDE models shows a remarkable compatibility with the Planck 2018 determination of $S_8 = 0.835 \pm 0.016$ within approximately $1.2\sigma$ for Model I and $1.3\sigma$ for Model II. This indicates that these interacting scenarios preserve the concordance with CMB measurements regarding the amplitude of matter fluctuations. However, our $S_8$ values remain higher than those obtained from some late-time-only analyses (e.g., RSD-only), suggesting that the inclusion of CMB priors plays a dominant role in anchoring the $S_8$ constraint.

Importantly, the interaction parameter $\epsilon$ was consistently constrained to be positive and small across all dataset combinations, implying a mild preference for a transfer of energy from dark energy to dark matter. This interaction introduces additional flexibility in the model, slightly increasing the best-fit values of both $H_0$ and $S_8$ compared to $\Lambda$CDM, although the shifts remain within $1\sigma$ uncertainties. Thus, while the IDE models provide a good fit to the data.

Finally, we emphasize that this analysis was performed within a simplified pseudo-Newtonian framework, suitable for the sub-horizon regime. A future implementation of these interactions in a relativistic Boltzmann code will be essential to obtain tighter constraints on $\epsilon$ and to perform a more definitive test of these models against the full CMB likelihood (Planck, ACT, SPT) and large-scale structure (LSS) information.
Furthermore, future high-precision data from next-generation surveys such as \textit{Euclid} \cite{Euclid:2025hny}, the Vera C. Rubin Observatory’s Legacy Survey of Space and Time (LSST) \cite{Zhan:2017uwu}, and the full DESI sample will provide unprecedented measurements of cosmic shear, galaxy clustering, and redshift-space distortions up to higher redshifts. These observations will significantly tighten constraints on the interaction parameter $\epsilon$ and offer a decisive test of whether IDE models can fully resolve the $S_8$ and $H_0$ tensions.

\begin{acknowledgments}
    AAE acknowledges financial support from School of Astronomy and Space Science, University of Science and Technology of China, Hefei, Anhui 230026, China. This study was financed in part by the Coordena\c{c}\~ao de Aperfei\c{c}oamento de Pessoal de N\'ivel Superior - Brasil (CAPES) - Finance Code 001. JFJ acknowledges financial support from  {Conselho Nacional de Desenvolvimento Cient\'ifico e Tecnol\'ogico} (CNPq) (No. 314028/2023-4). R.C.N. thanks the financial support from the CNPq under the project No. 304306/2022-3, and the Fundação de Amparo à Pesquisa do Estado do RS (FAPERGS, Research Support Foundation of the State of RS) for partial financial support under the project No. 23/2551-0000848-3.
\end{acknowledgments}

\appendix

\section{\label{APA}Constraints on the alternative parameterization: \texorpdfstring{$Q_{\text{II}}=\varepsilon H\Bar{\rho}_{\Lambda(t)}$}{Q=eps*H*rho\_L}}

We will now analyze the dynamic evolution of a flat $\Lambda(t)$CDM cosmological model with a phenomenological interaction term, $Q$, given by:
\begin{align}
    Q_{\text{II}} = \varepsilon  H\Bar\rho_{\Lambda(t)}\,. 
\end{align}
Switching from the time variable to the scale factor using $d/dt = aH\,d/da$, the continuity equations \eqref{DMQ} and \eqref{DEQ} are written as:
\begin{align}\label{eqrhodmp01}
     a\dfrac{\partial \Bar\rho_{\text{DM}} }{\partial a}+3\Bar\rho_{\text{DM}} &= \varepsilon \Bar\rho_{\Lambda(t)}\,,\\\label{eqrhop01}
     a\dfrac{\partial \Bar\rho_{\Lambda(t)} }{\partial a} &=-\varepsilon \Bar\rho_{\Lambda(t)}\,.
\end{align}
The differential equation for the energy density $\Bar\rho_{\Lambda(t)}$ expressed in \eqref{eqrhop01} has a solution given by:
\begin{align}\label{rhodepn0}
    \Bar\rho_{\Lambda(t)} &= \Bar\rho_{\Lambda 0}a^{-\varepsilon}\,,
\end{align}
For \eqref{eqrhodmp01}, we obtain:
\begin{align}
    a\dfrac{\partial \Bar\rho_{\text{DM}} }{\partial a}+3\Bar\rho_{\text{DM}} &= \varepsilon \Bar\rho_{\Lambda 0}a^{-\varepsilon}\,,
\end{align}
The solution to the above PDE is given by:
\begin{align}\label{rhodmpn0}
    \Bar\rho_{\text{DM}}(a)=\frac{a^{- {\varepsilon}-3}
   \left[a^{ {\varepsilon}} \left(\left( {\varepsilon}-3\right)\Bar\rho_{\text{DM}0}+ {\varepsilon}\Bar\rho_{\Lambda0}\right)-a^3{\varepsilon}\Bar\rho_{\Lambda0}\right]}{ {\varepsilon}-3}\,.
\end{align}

In this context, using the energy density \eqref{rhodmpn0} for $\Bar\rho_{\text{DM}}$ and the energy density \(\Bar\rho_{\Lambda}\) given in \eqref{rhodepn0}, the Friedmann equation is given by:
\begin{align}
H(a)^2=\frac{8 \pi  G}{3\left( {\varepsilon}-3\right)}\left\{a^{-3}\left[( {\varepsilon}-3)\left(\bar\rho_{\text{DM}0}+\bar\rho_{b0}\right)+ {\varepsilon}\bar\rho_{\Lambda0}\right]-3 a^{-\varepsilon} \bar\rho_{\Lambda0}\right\}.
\end{align}

As in the previous section, it is more advantageous for numerical analysis to use the parameter $\Omega_{M0} = \Omega_{DM0} + \Omega_{b0}$, since the data sample used in the statistical analyses does not constrain the baryon density and, since they are low-redshift data, the radiation term was omitted. For the interaction term $Q=\varepsilon  H\bar\rho_{\Lambda(t)}$, the equation \eqref{eqdelta} takes the form:
\begin{equation}\label{eqdeltan_final_N}
\small
    \begin{split}
        \delta''(a) = & -\frac{\delta'(a)}{2a \mathcal{P} \mathcal{N}} \left[ 
            \begin{split}
             & -9 a^7 \varepsilon (\varepsilon - 4) \Omega_{\Lambda 0}^2 \\
             & + a^{2\varepsilon} J \left( 3a L + 2(\varepsilon - 3)\Omega_{r0} \right) \\
             & + a^{3+\varepsilon} \Omega_{\Lambda 0} \Big( a(\varepsilon - 3) P_{mix} + 2 a \varepsilon \pi_{\Lambda} + 2\varepsilon(\varepsilon - 4)(\varepsilon - 3)\Omega_{r0} \Big)
            \end{split}
        \right] \\[10pt]
        & + \frac{\delta(a)}{2a \mathcal{P}^2 \mathcal{N}} \left[ 
            \begin{split}
            & 3 a^9 \varepsilon^2 (\varepsilon - 6)(\varepsilon - 2)\Omega_{\Lambda 0}^3 + 3 a^{3\varepsilon} J^3 \\
            & - a^{6+\varepsilon} \varepsilon \Omega_{\Lambda 0}^2 \Big( -(\varepsilon-3)^2 \varepsilon \Omega_{b0} + (\varepsilon-6)(\varepsilon-3)(8\varepsilon-15)\Omega_{dm0} + \varepsilon(\varepsilon-6)(8\varepsilon-15)\Omega_{\Lambda 0} \Big) \\
            & + a^{2+2\varepsilon} \varepsilon \Omega_{\Lambda 0} J \Big( a(\varepsilon-3)^2 (2\varepsilon - 7)\Omega_{b0} + a(2\varepsilon^2 - 13\varepsilon + 12)J + 2(\varepsilon-3)^3 \Omega_{r0} \Big)
            \end{split}
        \right]\,,
    \end{split}
\end{equation}
where the coupling terms $J$ (dark matter only) and $L$ (total matter) are defined as:
\begin{align}
    J &\equiv (\varepsilon - 3)\Omega_{dm0} + \varepsilon \Omega_{\Lambda 0}\,, \\
    L &\equiv (\varepsilon - 3)(\Omega_{b0} + \Omega_{dm0}) + \varepsilon \Omega_{\Lambda 0}\,.
\end{align}
The denominators $\mathcal{P}$ and $\mathcal{N}$ are given by:
\begin{align}
    \mathcal{P} &\equiv a^{\varepsilon} J - a^3 \varepsilon \Omega_{\Lambda 0}\,, \\
    \mathcal{N} &\equiv a^{\varepsilon} \left[ a L + (\varepsilon - 3)\Omega_{r0} \right] - 3 a^4 \Omega_{\Lambda 0}\,.
\end{align}
The auxiliary polynomials for the mixed terms are:
\begin{align}
    P_{mix} &= \varepsilon(2\varepsilon - 9)\Omega_{b0} + 2(\varepsilon^2 - 3\varepsilon - 9)\Omega_{dm0}\,, \\
    \pi_{\Lambda} &= (\varepsilon^2 - 3\varepsilon - 9)\Omega_{\Lambda 0}\,.
\end{align}

%\subsection{\label{ResII}Model II Analysis}

The observational constraints for {Model II} are summarized in Table~\ref{tab:param_model2}. 
We adopt the same priors described in Section~\ref{Res}. 
For the full dataset combination (RSD+CCs+PCMB+BAO+PP), we obtain
\( H_0 = 68.49 \pm 0.29 \) km s\(^{-1}\) Mpc\(^{-1}\),
\( \Omega_{\rm DM0} = 0.2542 \pm 0.0035 \), and
\( S_8 = 0.872 \pm 0.026 \).
The interaction parameter is constrained to
\( \varepsilon = 0.0155 \pm 0.0073 \), indicating a mild interaction compatible with zero at the
\( 2\sigma \) level.

The inferred value of the Hubble constant in Model II is robust against the presence of the interaction.
The full data combination yields
\( H_0 = 68.49 \pm 0.29 \) km s\(^{-1}\) Mpc\(^{-1}\),
which is essentially identical to that obtained for Model I
(\( H_0 = 68.51 \) km s\(^{-1}\) Mpc\(^{-1}\)) and fully consistent with the Planck 2018
\(\Lambda\)CDM results.
Similarly, the matter density parameters closely follow those of Model I, with
\( \Omega_{\rm DM0} = 0.2542 \pm 0.0035 \) and
\( \Omega_{b0} \simeq 0.0476 \) in the joint analysis.
This consistency demonstrates that, for small interaction strengths, the specific choice of the
interaction kernel does not significantly alter the background expansion history.

% --- TABELA DE PARÂMETROS (MODELO II) ---
\begin{table}[!htbp]
\centering
\footnotesize
\setlength{\tabcolsep}{4pt}
\begin{tabular}{lccccc}
\hline\hline
\textbf{Model/Dataset} & \textbf{$H_0$} & \textbf{$\Omega_{\text{DM}0}$} & \textbf{$\Omega_{b0}$} & \textbf{$\varepsilon$} & \textbf{$S_8$} \\
\hline
\multicolumn{6}{l}{\textbf{Model II}} \\
\hline
\quad PCMB+BAO & $68.63\pm 0.30$ & $0.2526\pm 0.0036$ & $0.04745\pm 0.00039$ & $0.0190\pm 0.0077$ & --- \\
\hline
\quad RSD+CCs+PP & $67.1\pm 3.2$ & $0.281\pm 0.031$ & $0.054^{+0.019}_{-0.034}$ & $0.0096^{+0.025}_{-0.031}$ & $0.897^{+0.046}_{-0.052}$ \\
\quad RSD+CCs+BAO+PP & $68.7\pm 3.3$ & $0.261^{+0.011}_{-0.012}$ & $0.0465^{+0.0046}_{-0.0035}$ & $0.009\pm 0.024$ & $0.868\pm 0.034$ \\
%\quad PCMB+CCs+BAO+PP & $68.52\pm 0.29$ & $0.2541\pm 0.0035$ & $0.04758\pm 0.00039$ & $0.0179\pm 0.0078$ & $0.90\pm 0.17$ \\
\quad RSD+CCs+PCMB+BAO+PP & $68.49\pm 0.29$ & $0.2542\pm 0.0035$ & $0.04766\pm 0.00038$ & $0.0155\pm 0.0073$ & $0.872\pm 0.026$ \\
\hline
\quad RSD+CCs+UNION3 & $65.4\pm 3.4$ & $0.307\pm 0.037$ & $0.054^{+0.019}_{-0.034}$ & $0.025^{+0.029}_{-0.041}$ & $0.919\pm 0.051$ \\
\quad RSD+CCs+BAO+UNION3 & $68.6\pm 3.3$ & $0.261\pm 0.012$ & $0.0464^{+0.0047}_{-0.0035}$ & $0.009^{+0.023}_{-0.026}$ & $0.867\pm 0.035$ \\
%\quad PCMB+CCs+BAO+UNION3 & $68.53\pm 0.30$ & $0.2538\pm 0.0035$ & $0.04757\pm 0.00039$ & $0.0185\pm 0.0078$ & $0.90\pm 0.17$ \\
\quad RSD+CCs+PCMB+BAO+UNION3 & $68.50\pm 0.30$ & $0.2539\pm 0.0035$ & $0.04766\pm 0.00038$ & $0.0160\pm 0.0074$ & $0.871\pm 0.026$ \\
\hline
\quad RSD+CCs+DESY5 & $65.8\pm 3.2$ & $0.301\pm 0.031$ & $0.054^{+0.019}_{-0.034}$ & $0.021^{+0.029}_{-0.035}$ & $0.915^{+0.045}_{-0.051}$ \\
\quad RSD+CCs+BAO+DESY5 & $68.1\pm 3.3$ & $0.270\pm 0.012$ & $0.0458^{+0.0047}_{-0.0035}$ & $0.019\pm 0.026$ & $0.878\pm 0.035$ \\
%\quad PCMB+CCs+BAO+DESY5 & $68.40\pm 0.29$ & $0.2553\pm 0.0035$ & $0.04771\pm 0.00039$ & $0.0181\pm 0.0078$ & $0.90\pm 0.17$ \\
\quad RSD+CCs+PCMB+BAO+DESY5 & $68.38\pm 0.29$ & $0.2553\pm 0.0035$ & $0.04778\pm 0.00038$ & $0.0158\pm 0.0074$ & $0.873\pm 0.026$ \\
\hline\hline
\end{tabular}
\caption{Constraints on free parameters for Model II with 68\% confidence limits.}
\label{tab:param_model2}
\end{table}
\begin{figure}[!htbp]
    \centering
    \includegraphics[width=0.5\linewidth]{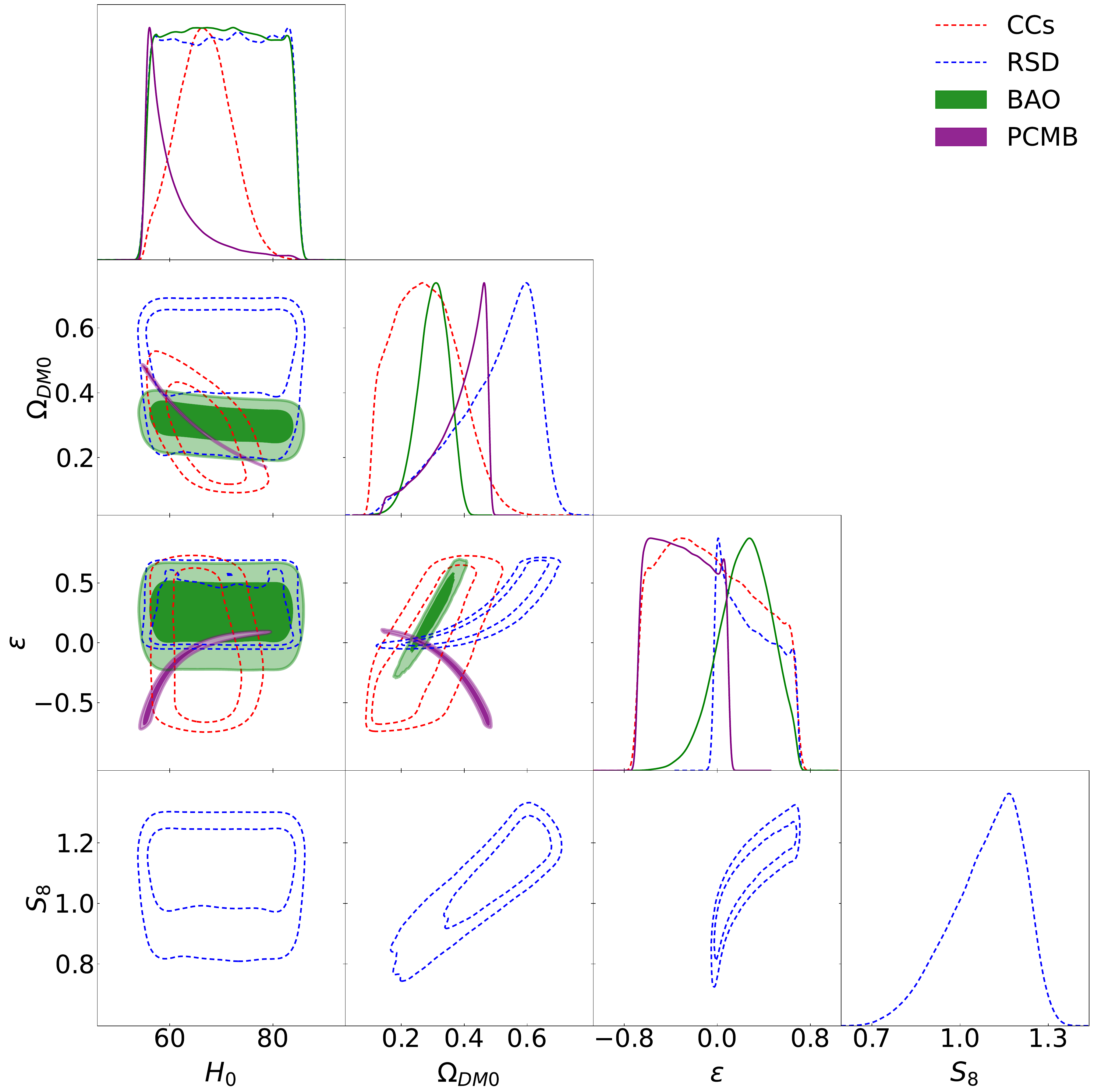}\includegraphics[width=0.5\linewidth]{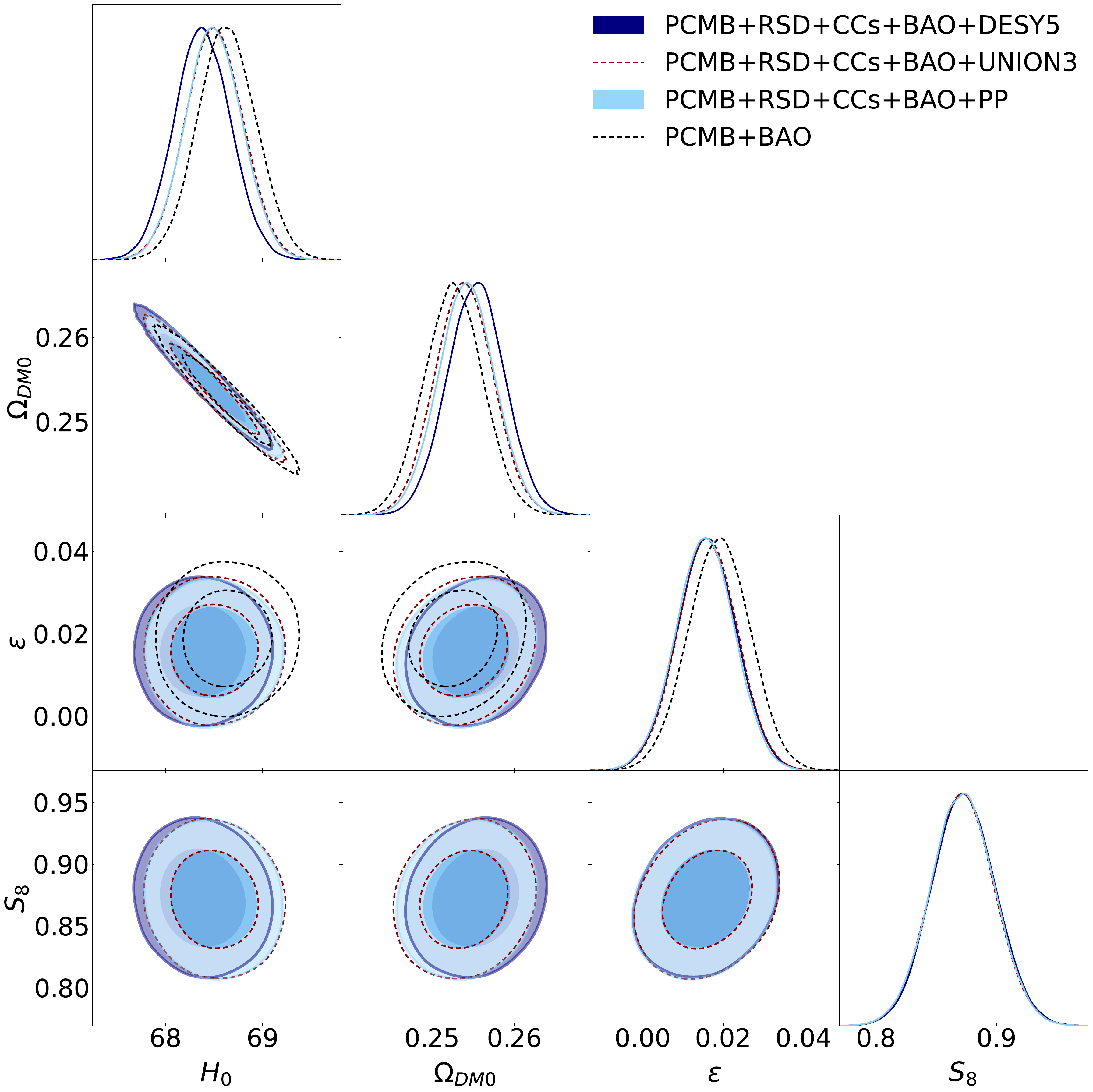}
    \caption{Marginalized posterior distributions and 68\% and 95\% C.L. contours for the main cosmological parameters $H_0$, $\Omega_{\rm DM0}$, $\varepsilon$, and $S_8$. The results are shown for different combinations of datasets, as indicated in the legend. \textbf{Left panel:} Constraints from individual datasets. \textbf{Right panel:} Constraints from the joint analysis of datasets.}
    \label{fig:contours_N0}
\end{figure}

Regarding the growth of structure, the combined datasets constrain the clustering amplitude to
\( S_8 = 0.872 \pm 0.026 \).
This value is compatible with the Planck 2018 result,
\( S_8 \simeq 0.834 \),
at approximately the \( 1.3\sigma \) level.
As in Model I, the inclusion of RSD data plays a crucial role in constraining \( S_8 \),
leading to a substantial reduction of uncertainties compared to background-only analyses.
The robustness of these constraints is further supported by the consistency between different
Type Ia supernova compilations, UNION3 and DESY5, as reported in Table~II, where the inferred
shifts in \( H_0 \) and \( S_8 \) remain well within statistical uncertainties.

A direct comparison between Model II, Model I, and the standard \(\Lambda\)CDM scenario shows that
Model II behaves almost indistinguishably from Model I.
Both interacting models yield slightly higher values of \( S_8 \) than the standard
\(\Lambda\)CDM result,
\( S_8 = 0.859 \pm 0.026 \) (for the same dataset combination),
while returning Hubble constant values that are statistically equivalent,
\( H_0 = 68.49 \) versus \( 68.47 \) km s\(^{-1}\) Mpc\(^{-1}\).
This comparison reinforces the interpretation of a small dark-sector interaction as a minimal
extension of \(\Lambda\)CDM: it introduces additional phenomenological freedom, encoded in
\( \varepsilon \), without compromising the tight constraints on the standard cosmological
parameters.
Consequently, the conclusions regarding the ability to accommodate combined datasets or alleviate
cosmological tensions drawn for Model I apply equally to Model II.

Figure~\ref{fig:contours_N0} shows the marginalized \( 1\sigma \) and \( 2\sigma \) confidence
contours for Model II.
The posterior distributions exhibit excellent consistency among individual datasets
(left panel), while the joint analysis (right panel) leads to significantly tighter constraints.
The usual geometrical degeneracies are observed, such as the negative correlation between
\( H_0 \) and \( \Omega_{\rm DM0} \).
In contrast, the interaction parameter \( \varepsilon \) displays no strong degeneracy with the
baseline cosmological parameters, further confirming the stability of the solution.

\bibliographystyle{apsrev4-1}
\bibliography{main_final}

\end{document}